# Research progress in high-pressure tuning of layered magnetic materials*

YUE Dongdong[1], GAO Xin[1], MU Congpu[1, 2], LIU Zhongyuan[1], ZHAI Kun[1]

1. Center for High Pressure Science, State Key Laboratory of Metastable Materials Science and Technology, Yanshan University, Qinhuangdao 066004, China

2. Hebei Key Laboratory of Microstructure Materials Physics, School of Science, Yanshan University, Qinhuangdao 066004, China

**Abstract**

Two-dimensional van der Waals materials have enormous potential applications in many fields due to their unique layered structure and excellent properties. Compared with three-dimensional bulk materials, layered systems are coupled by weak van der Waals interactions between layers, endowing them with much higher structural compressibility, particularly along the interlayer direction, which is more sensitive to external pressure. High pressure can expand the accessible phase space, enabling the synthesis of new materials or the retention of metastable phases. On the microscopic level, pressure can significantly tune the interlayer structure and interactions, induce changes in the electronic structure, and consequently give rise to a variety of rich physical properties. This article systematically introduces *in situ* high-pressure experimental approaches, including diamond anvil cells (DACs) combined with X-ray and spectroscopic techniques, high-pressure magnetic transport measurements, and emerging NV-center quantum sensing. It further reviews representative pressure-induced phenomena and underlying tuning mechanisms in layered magnetic materials, such as high-spin to low-spin transitions of transition-metal ions and the accompanying structural phase transitions and superconductivity; substantial enhancement of the Curie temperature ($T_c$) and continuous switching of magnetocrystalline anisotropy; and antiferromagnetic-to-ferromagnetic transitions achieved by modulating exchange interactions or via stacking engineering. Finally, we discuss future directions, including synergistic multi-field control by combining pressure with electric fields and twisted heterostructures, as well as strategies such as pressure quenching to retain high-pressure metastable magnetic phases at ambient conditions. These advances are expected to open new avenues for discovering novel magnetic states and developing high-performance device materials.

# 1. Introduction

High-pressure technology represents an interdisciplinary field that integrates deeply with earth sciences, materials science, and condensed matter physics. When materials are subjected to high-pressure conditions far exceeding ambient environments, reaching magnitudes of millions of atmospheres, their crystal and electronic structures often undergo significant changes. These alterations lead to novel physicochemical phenomena.

High-pressure techniques enable entirely new phase transition pathways, as illustrated in Figure 1. Consequently, researchers can synthesize new materials that are difficult to stabilize under ambient pressure, such as metallic hydrogen, unconventional superconductors, strongly correlated new phases. Thus, the high-pressure techniques had been approached serves as a vital strategy for exploring new materials[1-10]. High-pressure science acts as a crucial "accelerator" for material discovery and property regulation. It also provides a unique platform for understanding the intrinsic correlations among structure, electronics, and many-body interactions.

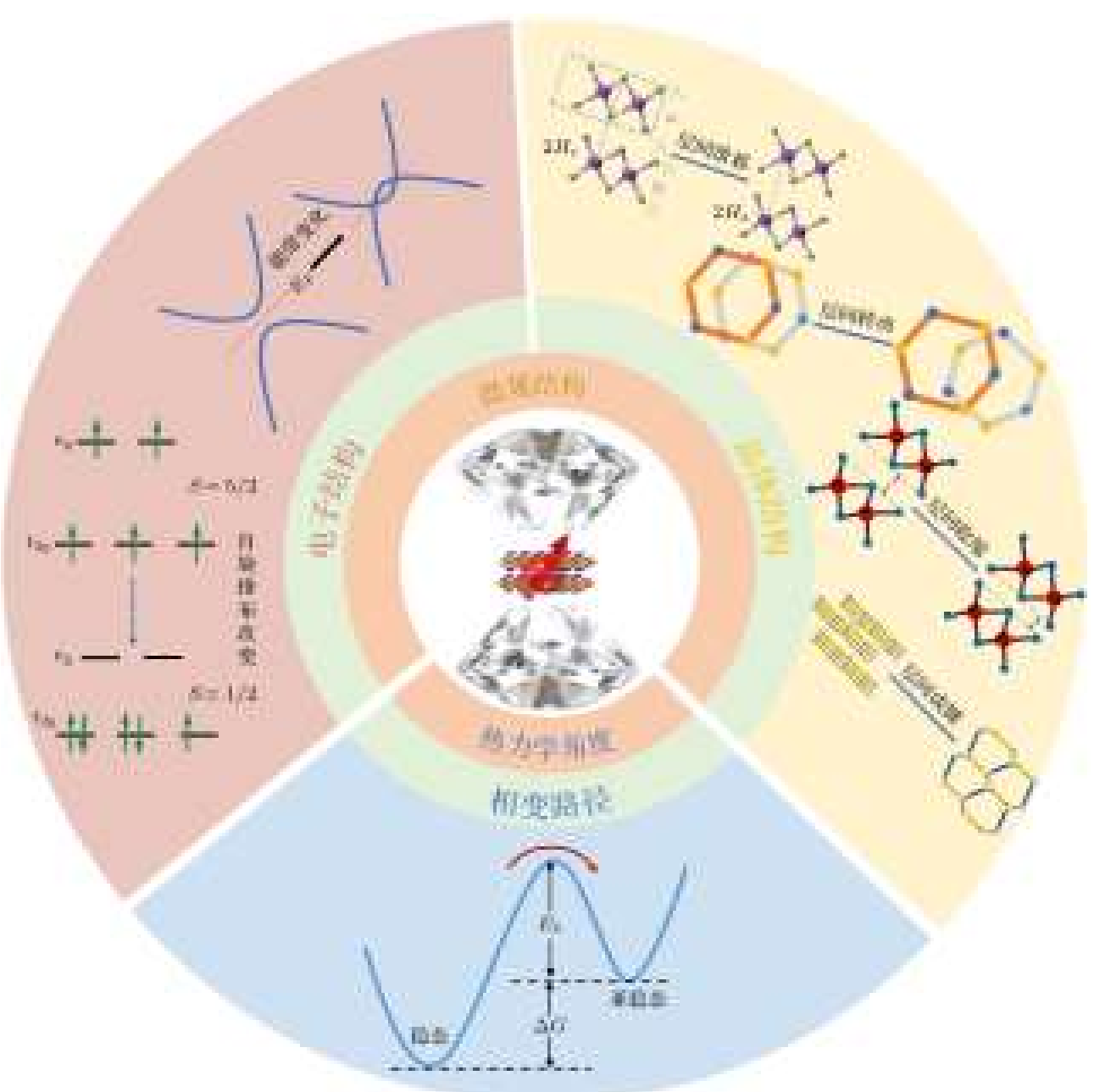

Fig. 1 High-pressure tuning effect on layered magnetic materials.

In recent years, materials (such as $CrI_3$, $Cr_2Ge_2Te_6$) have been found to maintain long-range magnetic order even in monolayer or few-layer. This discovery has sparked a surge of research into van der Waals magnetic materials[11,12]. The advantages of these van der Waals magnetic materials include: 1) the ability to retain long-range magnetic order at single-atom-layer thickness, offering potential for ultra-high-density storage applications[13]; 2) the weak van der Waals forces between layers allow for the construction of heterostructures without strict lattice matching constraints, providing greater possibilities for integrating spintronic devices[14-16]; and 3) the combination of weak interlayer van der Waals bonding and atomic-scale thickness makes them highly tunable to external stimuli. This facilitates the modulation of microscopic interactions, such as magnetic exchange and magnetocrystalline anisotropy via external fields (e.g., electric fields, strain, pressure, and chemical doping), thereby altering the macroscopic properties of materials[17-20].

High pressure serves as a crucial method for structural modulation and exhibits distinct advantages in layered van der Waals magnetic materials. Compared to three-dimensional bulk magnetic materials, layered systems possess greater compressibility due to weak interlayer van der Waals connections. They are particularly sensitive to pressure along the interlayer direction. Figure 1 summarizes the modulatory effects of high pressure on layered magnetic materials. As shown in Figure 1, external pressure readily triggers reconstruction processes involving interlayer degrees of freedom, such as compression of interlayer spacing, relative interlayer sliding, changes in stacking modes[21-25]. These structural changes significantly alter interlayer exchange pathways and anisotropy energy. From the perspective of electronic structure modulation, itinerant layered ferromagnetic systems (such as $Fe_xGeTe_2$ and $Fe_3GaTe_2$) exhibit magnetic orders closely related to the density of states near the Fermi level, band dispersion, and Fermi surface topology. Consequently, pressure induces lattice compression, enhances orbital overlap and interlayer coupling, and directly causes band renormalization and Fermi surface reconstruction. These changes ultimately modify magnetic exchange interactions and macroscopic magnetic properties[26].

Given the significant application potential of two-dimensional magnetic materials in spintronic devices and the rich physical property variations of layered magnetic materials under high pressure, this manuscript reviews recent progress in the high-pressure modulation of layered magnetic materials. We systematically introduce two aspects: the application of high-pressure experimental techniques in magnetic material research, and the typical phenomena and mechanisms of high-pressure modulation on the physical properties of layered magnetic materials. This review aims to demonstrate the unique advantages and broad prospects of high pressure as a key modulation tool. Specifically, it highlights its role in exploring new magnetic phases, enhancing key magnetic performance parameters, and advancing the development of two-dimensional magnetic materials and devices.

# 2. High-pressure experimental techniques applied to magnetic material research

The development of high-pressure science is inextricably linked to advancements in experimental techniques, which require specialized and sophisticated methods. High-pressure techniques include dynamic and static high-pressure methods. Static high-pressure techniques are most closely integrated with the study of material physical properties. In situ measurement techniques primarily involve large-volume presses and diamond anvil cell (DAC) technology[7-10]. Among these, in situ measurements based on DAC technology are the most widely applied. We next introduce the main high-pressure in situ experimental techniques relevant to magnetic measurements.

## 2.1 DAC

The DAC apparatus (shown in Figure 2(a)) is one of the most important inventions in the history of high-pressure science. This advance not only greatly extended the achievable pressure range, but more critically, also spurred major progress in DAC-based in situ measurement techniques for physical properties under high pressure[10]. The DAC utilizes two opposing diamonds to generate pressure. A metal gasket (pre-indented and drilled) placed between the two diamond anvils forms a sample chamber, where the sample is located in a confined space. The average pressure within the chamber (referred to as "pressure" in the high-pressure field) is defined as $P = F/A$, where $F$ is the force applied by the external loading mechanism and $A$ is the area of the diamond anvil face. The ultra-hard nature and spectral transparency of diamond provide a broad research platform for in situ physical measurements while achieving ultra-high pressures. Various new technologies based on DACs have emerged, including X-ray powder diffraction[27], X-ray single-crystal diffraction[28], metal gasket and liquid pressure-transmitting medium techniques[29], ruby pressure calibration[30], and the integration of DACs with combined extreme conditions such as low temperatures and strong magnetic fields[31]. These advancements have laid the foundation for modern high-pressure experimental techniques.

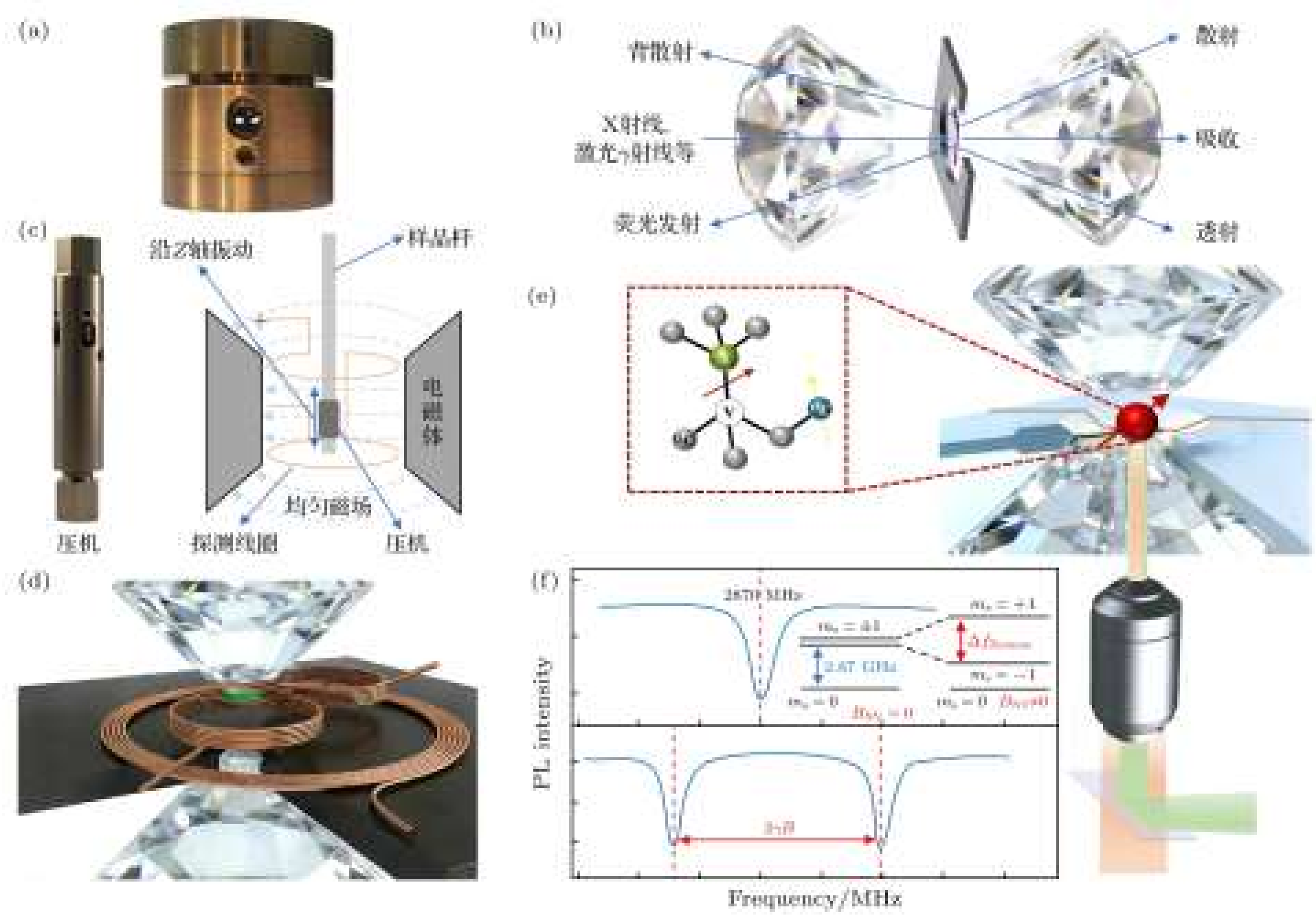

Fig. 2 (a) Photograph of a diamond anvil cell. (b) Interactions between different radiation sources and materials sealed in the diamond anvil chamber. Absorption techniques include XANES, EXAFS, and XMCD, while scattering techniques include diffraction and resonant inelastic X-ray scattering, among others. (c) Schematic illustration of direct-current magnetization measurements under high pressure. (d) Schematic illustration of alternating-current magnetization measurements under high pressure. (e) Schematic of the ODMR measurement setup under high pressure. (f) Typical resonance spectra measured under zero magnetic field and applied magnetic field, where the inset illustrates the schematic modulation effect of zero magnetic field and applied magnetic field on the energy levels.

## 2.2 Non-contact Measurements Based on Spectroscopy and X-rays

Using photons (such as visible light, X-ray photons, and γ-rays) as probes allows interaction with samples through diamond anvils. This interaction generates signals including transmission, scattering, diffraction, and fluorescence emission, as illustrated in Figure 2(b). These signals reflect the lattice structure, electronic structure, and magnetic properties of materials under high pressure. Non-contact measurements combined with DACs offer three distinct advantages. First, high-brightness beamlines and micro-beam spots enable measurements on extremely small samples, significantly improving the collection efficiency of weak signals[32]. Second, these techniques provide access to spin structure information. Raman spectroscopy identifies structural phase transitions and phonon-spin coupling. X-ray absorption near-edge structure (XANES) is sensitive to valence states, coordination environments, and changes in electronic structure. It is used to study pressure-induced charge transfer, spin-state transitions, and coordination reconstruction[33]. Extended X-ray absorption fine structure (EXAFS) quantitatively determines nearest-neighbor atomic distances, coordination numbers, and disorder. This makes it suitable for investigating local structural evolution before and after phase transitions[34]. X-ray emission spectroscopy (XES) is commonly used to obtain information on changes in spin states and local magnetic moments. For instance, the $K_\beta$ emission line shape is sensitive to the spin states of 3d transition metals. This technique detects phenomena such as high-spin to low-spin transitions under high pressure[35]. X-ray magnetic circular dichroism (XMCD) utilizes the difference in absorption of circularly polarized X-rays in magnetized samples. It achieves element-resolved magnetic detection and measures the spin and orbital magnetic moments of specific elements under high pressure[36]. Third, DACs can be integrated with various external field environments. These include cryogenic temperatures (down to the mK range), high temperatures (laser heating up to thousands of K), strong magnetic fields (superconducting or pulsed magnets up to tens of tesla), and electric fields. This integration constructs a multi-parameter platform for measurements under extreme conditions.

## 2.3 Magnetic Measurements under High-Pressure

The total magnetic moment is an extensive quantity, with signal amplitude approximately proportional to sample volume or mass. Under ambient conditions, superconducting quantum interference device (SQUID) magnetometers and vibrating sample magnetometers (VSM) achieve high-sensitivity magnetic measurements ($10^{-7}$-$10^{-9}$ emu). However, in high-pressure experiments, samples are confined within piston-cylinder cells or the tiny chambers of DACs. Sample mass can be as low as the microgram level. Furthermore, the high-pressure apparatus itself (including CuBe, NiCrAl, Re gaskets, diamonds, and sealants) introduces significant background signals and parasitic responses. Materials in the pressure apparatus may exhibit paramagnetism, diamagnetism, or contain trace ferromagnetic impurities. These factors pose challenges for macroscopic magnetic measurements. Current primary measurement methods include DC methods, AC methods, and nitrogen-vacancy (NV) center methods.

1) DC method. A sample of known mass is typically loaded into a non-magnetic high-pressure piston-cylinder or diamond anvil apparatus. The sample and apparatus are measured together within detection coils. The background signal of the apparatus is then subtracted to isolate the magnetic signal of samples[37-39], as shown in Figure 2(c).

2) AC method. This method typically employs mutual inductance coils. An excitation coil generates an AC magnetic field $H_{ac}$ at frequency $f$. The magnetic response of the sample causes changes in magnetic flux. A detection coil measures the induced voltage. A lock-in amplifier extracts the real part (corresponding to $\chi'$) and the imaginary part (corresponding to $\chi''$), as shown in Figure 2(d). Because the coils are positioned closer to the sample, this method directly detects flux changes caused by variations in sample magnetism[40-42].

3) NV center method. In recent years, color-center quantum sensing, represented by diamond nitrogen-vacancy (NV) centers, has provided a distinct alternative to traditional magnetic characterization for micro-area magnetic measurements. The optical transitions of NV centers are sensitive to spin states. Optical methods achieve spin polarization and readout. Optically detected magnetic resonance (ODMR) maps local magnetic fields (including static fields, AC fields, and magnetic noise) into quantifiable spectral line shifts and broadening. This enables nanoscale magnetic imaging and point measurements[43-48]. Magnetic detection using NV center sensors typically involves microwave excitation and fluorescence collection. In the absence of an external magnetic field, the NV center ground state is a spin triplet. Zero-field splitting exists between the $m_s$ = 0 and $m_s$ = ±1 states. The $m_s$ = ±1 states are nearly degenerate, with a corresponding resonance frequency of approximately 2.87 GHz. Upon excitation with short-wavelength laser light (e.g., 532 nm), the system emits fluorescence via radiative transitions. It may also undergo non-radiative transitions through a metastable singlet channel. This non-radiative channel has a higher probability for the $m_s$ = ±1 states, resulting in lower fluorescence intensity for these states compared to the $m_s$ = 0 state. Consequently, applying microwaves near 2.87 GHz drives resonant transitions from $m_s$ = 0 to $m_s$ = ±1. The resulting population transfer to the $m_s$ = ±1 states cause an observable decrease in fluorescence intensity, forming the characteristic ODMR spectrum (as shown in Figure 2(f)). In the presence of an external magnetic field or stray fields from magnetic samples, the Zeeman effect splits the $m_s$ = ±1 states. Their resonance frequencies shift approximately linearly with the magnetic field component along the NV axis. The frequency difference between the $m_s$ = +1 and $m_s$ = -1 resonance peaks is approximately $2\gamma_e B$. Under high-pressure conditions, NV centers with different orientations experience varying local stresses. This alters their spin energy level structures, leading to more complex changes in ODMR spectral lines, such as splitting, weakening, and broadening. Experimentally, local magnetic fields can be derived from ODMR peak positions and splitting magnitudes. These data are further used to analyze magnetization distribution in samples under high pressure and diamagnetic responses in superconductors[49-51].

## 2.4 Transport Experimental Techniques under High-Pressure

Direct magnetic measurements under high pressure using the aforementioned DC and AC methods often introduce large background signals, making it difficult to obtain accurate magnetic information. Although the NV center method offers extremely high sensitivity, it presents certain technical barriers. Consequently, high-pressure magnetic transport measurements have become a primary method for studying layered magnetic materials. These measurements explore changes in magnetism and electronic structure by measuring magnetoresistance and Hall resistance under high pressure. Interpreting anomalous Hall effect data for layered magnetic materials under high pressure is a significant research focus. To address the long-standing challenge of quantitatively measuring the susceptibility of micro-scale ferromagnets, Qin et al.[52] proposed a quantitative analysis method based on the anomalous Hall effect (AHE) and Arrott plots. This method allows for the in-situ extraction of susceptibility and identification of magnetic anisotropy orientation on individual micrometer-scale van der Waals ferromagnetic nanoflakes.

### 2.4.1 Micro/Nano-electrode Integration and Sample Transfer

From a technical perspective, nanoflakes of layered materials (few-layer to single-layer) are smaller in size than bulk samples. Manual wiring schemes used for traditional bulk materials are more difficult to apply. Micro-nano fabrication techniques are typically required to integrate electrodes onto diamond anvil surfaces. Nanometer-thick samples are then directionally transferred onto these electrodes for electrical measurements[53-56], as shown in Figures 3(a) and (b). Micro-nano electrode integration technology provides an important technical foundation for conducting high-pressure transport measurements on few-layer samples.

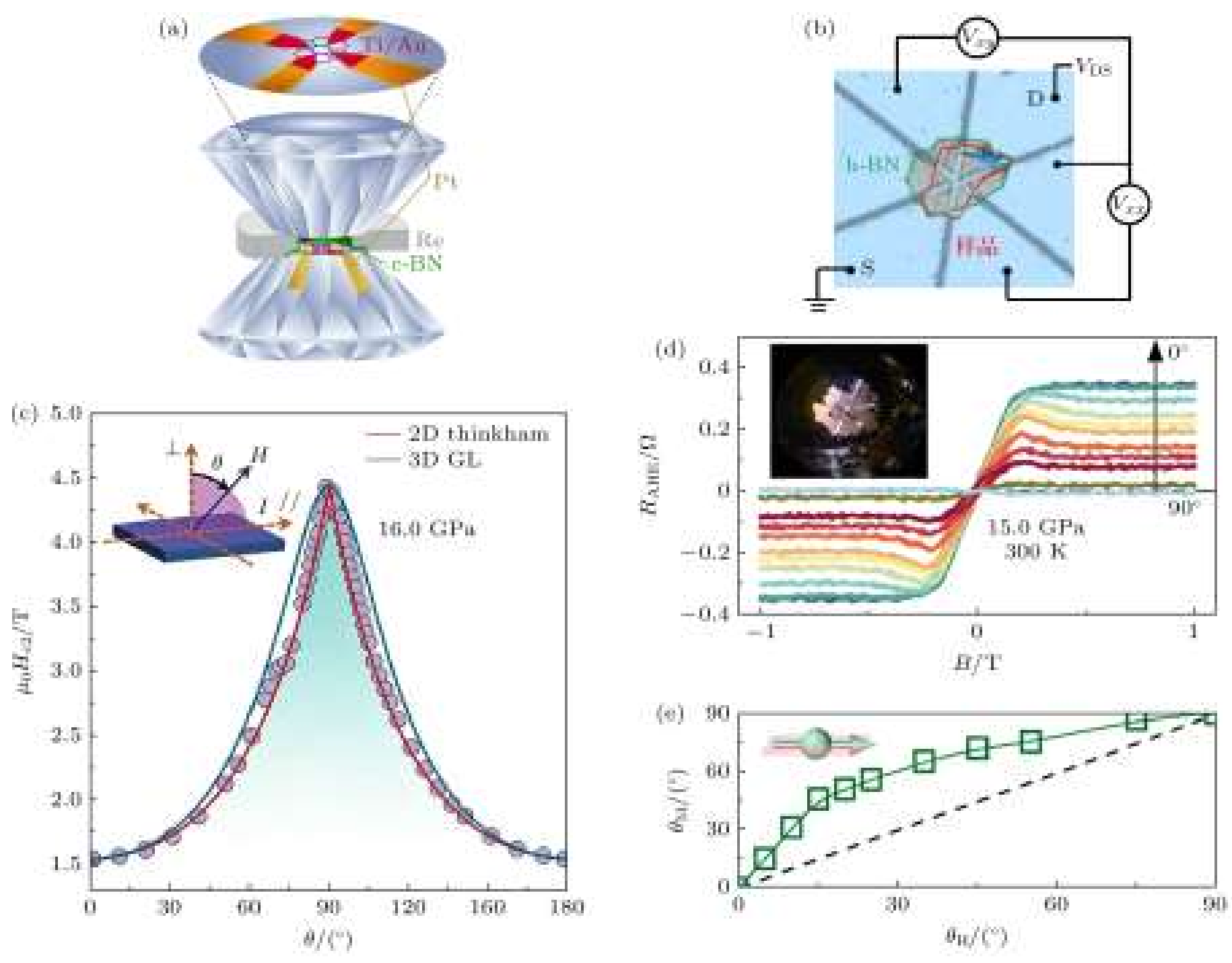

Fig. 3 (a) Schematic illustration of electrodes integrated on the diamond culet[55]; (b) optical image of transferring a layered sample onto electrodes fabricated on the diamond culet[56]; (c) angular dependence of the upper critical field of superconductivity under high pressure[53]; (d) angular dependence of the Hall resistance; (e) analysis of the experimental results using the Stoner-Wohlfarth single-domain model; the $\theta_M$-$\theta_H$ curve falls in the upper-left region of the diagram, indicating in-plane magnetic anisotropy[61].

## 2.4.2 In-Situ High-Pressure Angular Transport Measurements

Anisotropy constitutes a critical aspect of material property research. Angular magnetotransport measurements can reveal superconducting dimensionality, magnetocrystalline anisotropy, Fermi surface geometry, and the origin of multiferroicity[53,57-61] (as shown in Fig. 3(c)—(e)). However, performing angle-dependent property measurements under high pressure (combined with strong magnetic fields and low temperatures) presents significant challenges. In large-bore low-temperature magnets, rotating devices can be placed at the bottom of the cryostat. Nevertheless, this approach faces obstacles such as limited sample chamber volume, restricted rotation of pressure cells, large angular deviations at low temperatures, and incompatibility with ambient-pressure sample rods. Recently, Yue et al.[62] improved traditional measurement schemes by adopting a modular design. They separated the mechanical rotating device from the sample rod and positioned it at the top of the sample chamber. Since this mechanical rotating part operates at room temperature, it achieves high-precision continuous rotation. Furthermore, the design incorporates various sample holders and utilizes transverse magnetic fields provided by split-pair magnets. This configuration enables both in-plane-to-in-plane and in-plane-to-out-of-plane angular rotations. These advancements have significantly propelled experimental technology and elucidated the magnetic properties of various layered materials under high pressure.

# 3. Typical phenomena of layered magnetic materials under high-pressure

## 3.1 Tuning of Spin States

High pressure can modulate the spin arrangement of transition metal ions (d electrons). Under high pressure, the compression of metal-ligand bond lengths significantly increases the crystal field splitting energy. When this energy exceeds the Hund's exchange energy, electrons tend to pair up in lower-energy orbitals, thereby driving a transition from high-spin to low-spin states. Wang et al.[63-65] observed spin crossover phenomena under high pressure in antiferromagnetic layered material systems such as $FeP\mathit{X}_3$ ($X$ = S, Se) and $MnP\mathit{X}_3$ ($X$ = S, Se). Taking $FeP\mathit{X}_3$ ($X$ = S, Se) as an example, X-ray emission spectroscopy measurements of Fe under high pressure show that the $K_{\beta'}$ peak disappears at specific pressures (as shown in Fig. 4(a)—(c)). This indicates that the d-electron configuration of Fe changes from a high-spin to a low-spin state. This high-spin to low-spin transition is associated with a first-order structural phase transition, manifested as a significant volume contraction (exceeding 10%) (as shown in Fig. 4(d) and (e)), accompanied by a semiconductor-to-metal transition. For $FePSe_3$, superconductivity emerges near the phase transition after the system enters a high-pressure phase characterized by a low-spin, non-magnetic, and metallic state. The onset $T_c$ is approximately 2.5 K at ~9 GPa and reaches about 5.5 K at ~30 GPa (as shown in Fig. 4(f)). This demonstrates that pressure-driven spin crossover serves as a viable pathway for discovering new superconducting phases in unconventional structural parents.

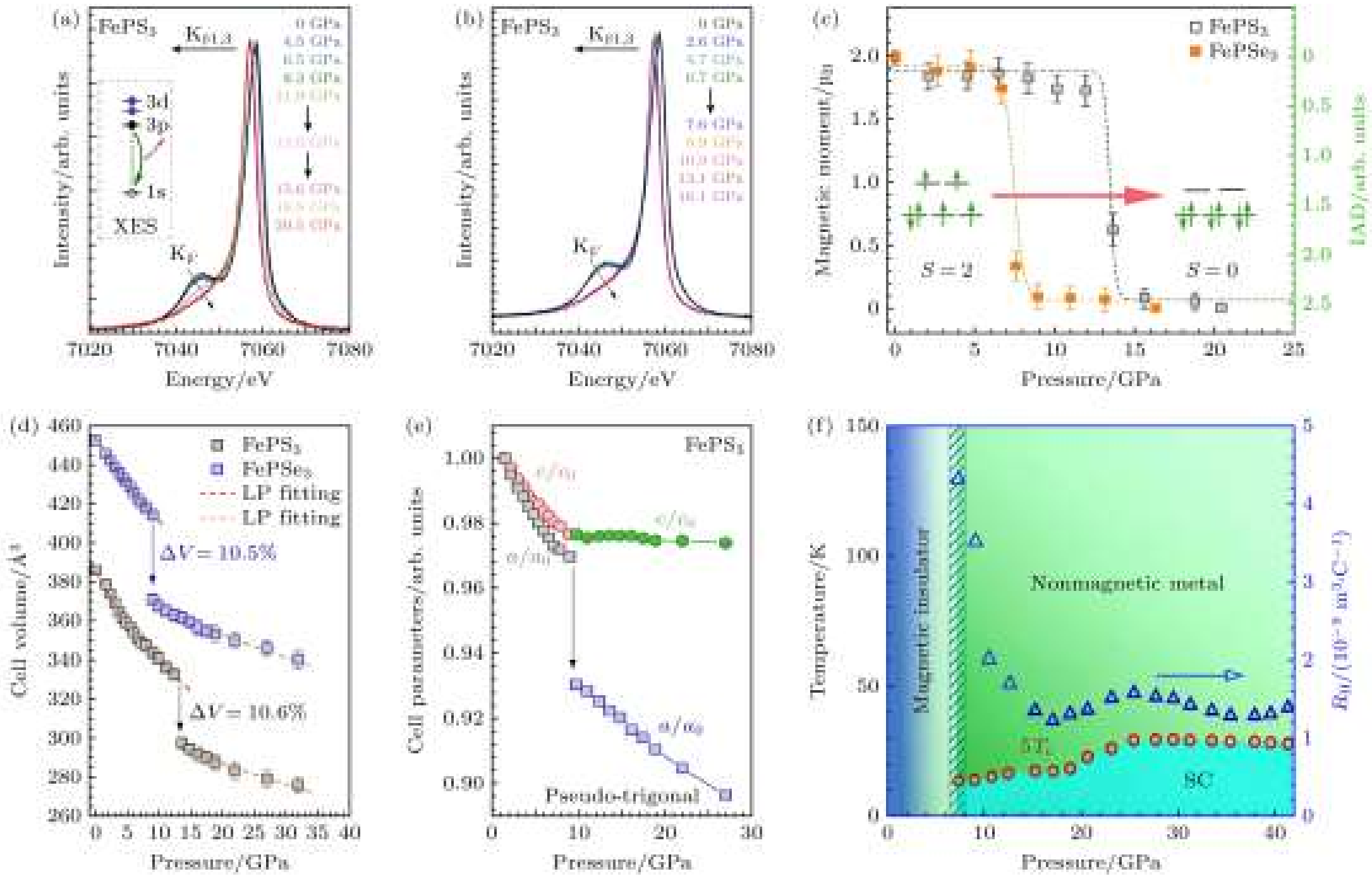

Fig. 4 (a) Fe $K_\beta$ X-ray emission spectra of $FePS_3$ and (b) $FePSe_3$ under different applied pressures; (c) pressure dependence of the magnetic moments of $FePS_3$ and $FePSe_3$ obtained from the integrals of the absolute values of the difference spectra analysis; (d) unit-cell volume as a function of pressure for $FePS_3$ and $FePSe_3$; (e) pressure dependence of the lattice parameters of $FePSe_3$; (f) temperature-pressure phase diagram of $FePSe_3$ [64].

## 3.2 Regulation of Curie Temperature and Magnetocrystalline Anisotropy

Beyond modulating local spin configurations, high pressure also regulates magnetic long-range order. For two typical layered magnetic materials, $Fe_5GeTe_2$ and $Fe_3GaTe_2$, their response to high pressure is primarily manifested in the synchronous tunability of the Curie temperature and magnetocrystalline anisotropy (as shown in Fig. 5(a) and (b)). In $Fe_3GaTe_2$, high pressure induces a distinct continuous transition in anisotropy. The system retains strong perpendicular magnetic anisotropy (PMA) in the low-pressure regime. At higher pressures, the easy magnetic axis undergoes a continuous rotation from out-of-plane to tilted, and finally to in-plane. Concurrently, $T_c$ exhibits a "dome-like" evolution with pressure. It gradually increases to approximately 480 K near 10.3

GPa, before declining at higher pressures[61]. For $Fe_5GeTe_2$, increasing pressure enables controllable evolution among different magnetocrystalline anisotropic ferromagnetic phases. The system transitions from an in-plane easy axis (FM1) to an out-of-plane easy axis (FM2), and finally to a nearly isotropic magnetic state (FM3). This transition is accompanied by a significant increase in the Curie temperature. Notably, the high-temperature ferromagnetic FM3 state emerging at higher pressures persists to elevated temperatures and remains stable after pressure release. This retention further elevates the Curie temperature of the material to above 400 K, demonstrating the trapping of a high-pressure-induced metastable high-temperature ferromagnetic state[66]. Regarding potential mechanisms, both materials share a common feature. High pressure simultaneously influences exchange interactions and magnetocrystalline anisotropy energy by compressing the lattice and reshaping the electronic structure. Pressure enhances orbital overlap and drives the reconstruction of the density of states or Fermi surface near the Fermi level, thereby altering magnetic exchange interactions and $T_c$. Meanwhile, the sensitivity of spin-orbit coupling to the crystal field environment causes significant changes in magnetocrystalline anisotropy energy under pressure, leading to the flipping of the easy magnetization axis or a trend toward isotropy.

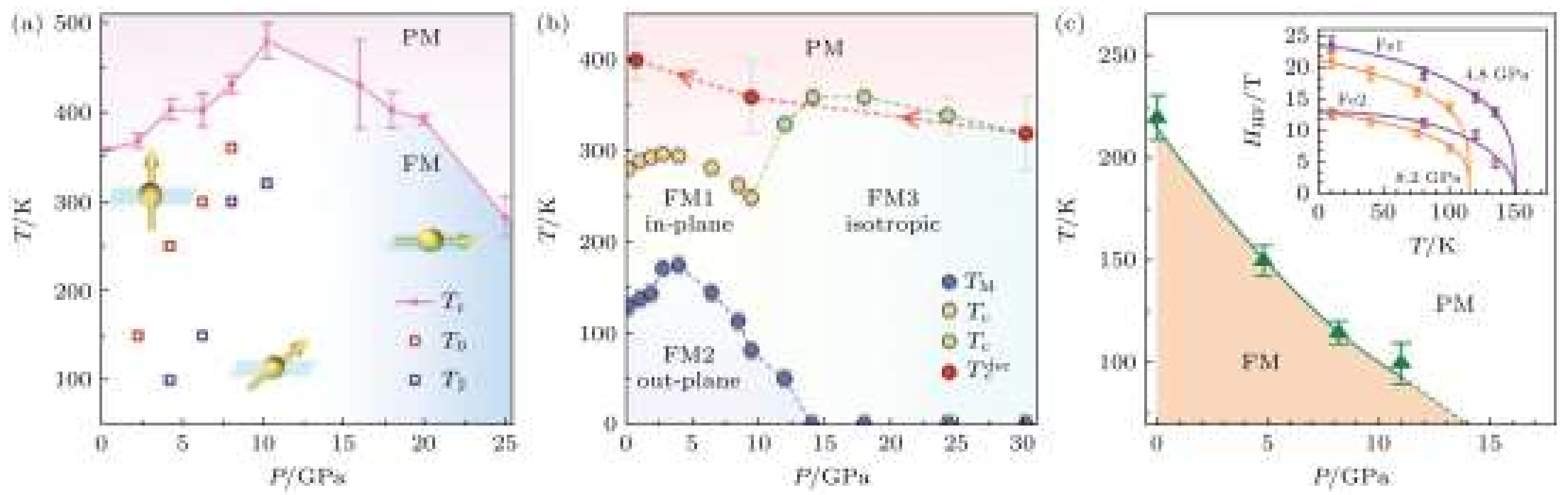

Fig. 5 (a) High-pressure phase diagram of $Fe_3GaTe_2$[61], (b) $Fe_5GeTe_2$[66], and (c) $Fe_3GeTe_2$[68]. The inset in panel (c) shows the temperature dependence of the hyperfine fields at the Fe1 and Fe2 sites in the $Fe_3GeTe_2$ sample.

$Fe_3GeTe_2$ ($T_c$ ~ 180—230 K), which shares a similar structure with $Fe_3GaTe_2$, exhibits distinct pressure response characteristics. As pressure increases, the $T_c$ of $Fe_3GeTe_2$ decreases in an approximately monotonic manner. The material enters a paramagnetic state near the critical pressure (~15 GPa)[67-69], as shown in Figure 5(c). These markedly different responses to pressure in $Fe_3GeTe_2$ and $Fe_3GaTe_2$ may hint at the microscopic mechanisms underlying their significant $T_c$ differences.

### 3.3 Regulation of Magnetic Ordered States

High pressure can directly alter the strength or even the sign of interlayer or intralayer exchange interactions by continuously compressing the interlayer spacing and modulating bond lengths and angles. This process induces magnetic order transitions. For instance, Co-doped $Fe_3GaTe_2$ can transition to an antiferromagnetic (AFM) ground state (Figure 6(a)). High-pressure regulation studies reveal that the sample reverts from the AFM state to a ferromagnetic (FM) state at 4.1 GPa. Furthermore, the magnetic anisotropy exhibits an out-of-plane easy axis characteristic (Figures 6(b)—(d)). When the pressure increases to 12.3 GPa, $T_c$ rises to a maximum of approximately 282 K. With further pressurization, $T_c$ begins to decline, presenting a "dome-shaped" phase diagram similar to that of the parent phase $Fe_3GaTe_2$ (Figure 6(e)). High-pressure X-ray diffraction (XRD) results indicate no signs of structural phase transition in the sample at 20 GPa. This suggests that the transition in ferromagnetic order results from pressure-induced changes in the electronic structure[56].

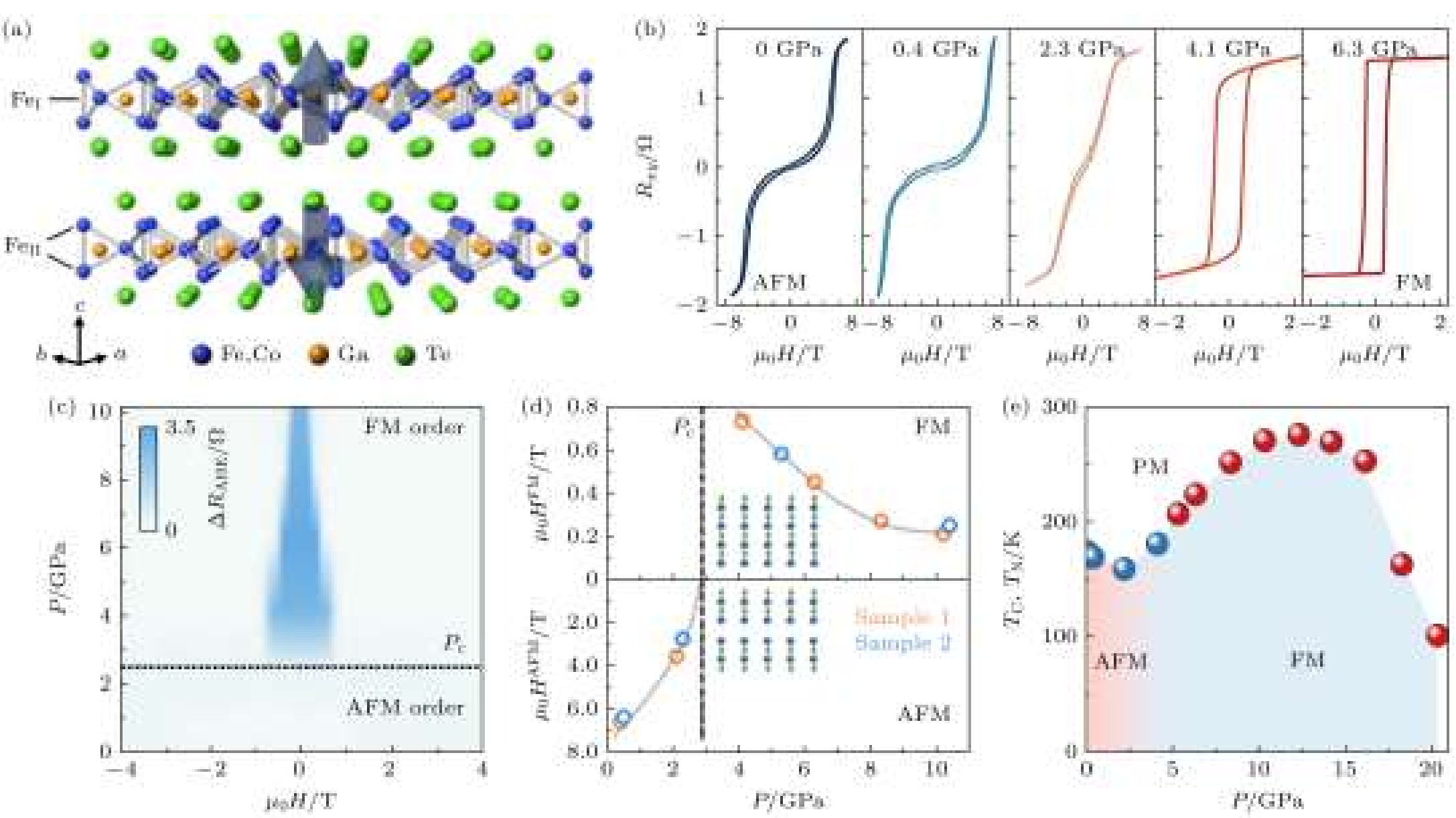

Fig. 6 (a) Crystal structure of $(Fe_{1-x}Co_x)_3GaTe_2$. The arrows schematically indicate the magnetic configurations. (b) Anomalous Hall resistance of a nanoflake of the layered magnet ($(Fe_{1-x}Co_x)_3GaTe_2$ ($x$ = 0.24) under different pressures. (c) Pressure- and magnetic-field-dependent Hall signals at 10 K measured under opposite magnetic-field sweeping directions. Here, $\Delta R_{AHE} = | R_{AHE}(H_{increase}) - R_{AHE}(H_{decrease}) |$. For out-of-plane ferromagnetic ordering, $\Delta R_{AHE}$ is nonzero. However, for antiferromagnetic ordering, $\Delta R_{AHE}$ is close to 0 because the magnetic hysteresis loop is negligible. Therefore, $P_c$ denotes the critical pressure at which the AFM-to-FM transition occurs. (d) Pressure dependence of the spin-flop field $H_c^{AFM}$ (and the coercive field $H_c^{FM}$ for FM ordering) at 10 K. (e) High-pressure phase diagram of $(Fe_{1-x}Co_x)_3GaTe_2$ ($x$ = 0.24)[56].

High pressure acts on interlayer degrees of freedom. In addition to compressing the interlayer spacing and altering bond lengths and angles, it may drive relative interlayer sliding and misalignment. These processes modify the original stacking structure and induce changes in physical properties, including alterations in magnetic ordered states. In the typical $CrI_3$ system, thin-layer samples exhibit a monoclinic stacking structure at ambient pressure, which corresponds to antiferromagnetic interlayer coupling. Under hydrostatic high pressure (~1.8 GPa), the system undergoes an irreversible transition from monoclinic to rhombohedral stacking. Consequently, the interlayer coupling shifts from antiferromagnetic to ferromagnetic, converting the sample into a ferromagnetic phase. This pressure-induced ferromagnetic state persists up to approximately 60 K, approaching the Curie temperature scale of bulk $CrI_3$[70,71]. These studies clearly reveal the influence of interlayer stacking on interlayer coupling in two-dimensional magnets, thereby modifying long-range magnetic order. This demonstrates that pressure not only modulates macroscopic magnetism by compressing interlayer spacing but also triggers stacking rearrangements that alter interlayer exchange interactions and induce new magnetic effects. These findings provide a pathway for using high pressure to achieve stacking engineering, twisted moiré magnetism, and reconfigurable spin devices.

# 4. Outlook: Future Directions for High-Pressure Modulation in Layered Magnetic Materials

The unique advantage of high pressure lies in its ability to alter atomic distances in a "clean" manner without introducing chemical disorder. Therefore, it is particularly suitable for exploring new magnetic phases, extreme magnetic states, and pressure-induced exchange mechanisms. However, high-pressure modulation has inherent limitations. First, high-pressure regulation of electronic structure often couples with changes in multiple structural degrees of freedom, such as bond lengths, bond angles, interlayer sliding, and stacking rearrangements. This makes its control parameters less "pure" than those of electric fields (e.g., ionic liquid gating). Second, in situ characterization and device integration under high pressure are complex. Furthermore, many pressure-induced phases revert to ambient-pressure structures upon decompression, making them difficult to utilize under ambient conditions. Because high pressure can drive systems into structural and band configurations inaccessible at ambient pressure, yet faces challenges regarding strong coupling between control parameters and the lattice, as well as the difficulty of retaining high-pressure phases, future research must focus on the following directions. One direction is to develop synergistic coupling between high pressure and other modulation methods to achieve precise multi-parameter control. Another is to address the stable retention of magnetic high-pressure phases at ambient pressure, thereby transforming new properties observed under high pressure into new materials and device platforms usable at ambient conditions.

Coupling high pressure with multiple modulation methods. As a research platform, high pressure can be combined with various modulation techniques to transition from single-parameter control to multi-parameter synergistic control. This allows for the exploration and discovery of novel magnetic properties in a broader parameter space. Recent studies demonstrate the powerful capabilities of comprehensive modulation. Chen et al.[72] constructed and measured field-effect transistors in situ within diamond anvil cells, achieving synergistic high-pressure and gate-voltage modulation of $MoS_2$. This significantly enhanced carrier mobility and contact performance. Yankowitz et al.[73] combined twisted bilayer graphene with high-pressure modulation. In devices with a twist angle of 1.27°, pressure induced a superconducting transition temperature ($T_c$) of approximately 3 K. This value is nearly an order of magnitude higher than the 0.3-0.4 K typically observed at the magic angle (1.1°), highlighting the powerful synergy between twist engineering and high-pressure modulation. Wang et al.[74] combined high pressure and gate voltage modulation (stacking + high pressure + gate voltage) in graphene/hexagonal boron nitride moiré superlattices. They demonstrated that pressure significantly enhances the moiré potential and observed the theoretically predicted "tertiary energy gap" for the first time. Therefore, high pressure is no longer merely a single-parameter control method. It serves as a comprehensive platform that can be deeply coupled with electric fields, heterogeneous stacking, or optical fields. High pressure drives the system into new lattice and band configurations, while other methods further fine-tune carrier concentration and nearest-neighbor interactions. This approach enables the construction of multidimensional phase diagrams and the realization of richer physical properties within specific high-pressure phases.

Retention of high-pressure phases. From the perspective of fundamental materials research, discovering new layered magnetic materials remains a fundamental issue. High pressure expands the phase space and provides phase transition pathways different from those at ambient pressure. This makes many structures and chemical configurations inaccessible at ambient pressure possible, offering broad opportunities for discovering new layered materials. This is particularly relevant for layered van der Waals materials, which often exhibit various stacking and faulting configurations. High pressure can reduce sliding barriers and stabilize specific stacking forms. If these forms can be retained after decompression, this will become an important pathway for discovering new materials.

Although in situ high-pressure measurements reveal rich structural and physical property evolutions, stably "retaining" these pressure-induced properties and structures at ambient pressure remains challenging. This constitutes a key direction for next-stage research. In recent years, several strategies have shown promising application prospects. For example, the group of Chu Jingwu[75,76] proposed pressure quenching (rapid decompression at specific pressure and temperature conditions). This method utilizes the kinetic energy barriers of phase transitions (such as the latent heat of first-order phase transitions or barriers from structural rearrangements) to retain pressure-induced metastable phases at ambient pressure. This approach successfully stabilized the high-pressure superconducting phase of materials like $Bi_{0.5}Sb_{1.5}Te_3$ at ambient pressure. Li et al.[77] retained the high-pressure metastable phase of the $Rb_6Re_6S_8I_8$ system at ambient pressure through repeated loading and unloading cycles. In this system, the $[Rb_6I_2]^{4+}$ framework is soft and plastic, while the $[Re_6S_8I_6]^{4-}$ clusters are rigid and elastic. This difference in rigidity allows external pressure to selectively amorphize the framework while maintaining the ordered state of the clusters. This structural characteristic "freezes" the pressure-induced metastable state to ambient pressure upon decompression, with photocurrent increasing as the number of compression-decompression cycles increases. He et al.[78] applied 5 GPa pressure at room temperature and introduced torsional deformation. Shear strain induced the transition from $\alpha$-Ti to $\beta$-Ti, and the $\beta$-Ti phase remained stable after decompression. Typically, the $\beta$ phase in pure titanium appears only at high temperatures (>1155 K) or under extremely high hydrostatic pressure (>52 GPa) at ambient pressure. These studies indicate that introducing additional operational dimensions, such as temperature variation, shear force, and rapid pressure increase/decrease, can significantly enhance the retainability of high-pressure metastable phases at ambient conditions. This provides important references and implementation pathways for retaining high-pressure phases of layered magnetic materials.

In summary, high pressure is a highly scalable method that combines physical property modulation, material synthesis, and the exploration of novel phases. Particularly in layered van der Waals systems, the unique interlayer modulation capabilities of pressure (such as interlayer spacing compression, enhanced interlayer coupling, and stacking reconstruction) provide pathways for magnetic modulation that are difficult to achieve with other external fields. High pressure will continue to play a key role in understanding mechanisms, improving performance, and discovering new phases and materials in layered magnetic materials. It is expected to drive further breakthroughs in material systems featuring high Curie temperatures, strong perpendicular anisotropy, and integrable spin devices.